\documentclass[submit,techrep,JIP]{ipsj}

\usepackage{url}

\usepackage[pdftex]{graphicx}

\usepackage{latexsym}

\def\Underline{\setbox0\hbox\bgroup\let\\\endUnderline}
\def\endUnderline{\vphantom{y}\egroup\smash{\underline{\box0}}\\}
\def\|{\verb|}

\makeatletter
\def\@oddhead{}%
\def\@evenhead{}%
\def\ps@IPSJTITLEheadings{}
\makeatother

\begin{document}

\title{Different Types of Voice User Interface Failures
\\ May Cause Different Degrees of Frustration
}
\affiliate{IPSJ}{Waseda University}


\author{Shiyoh Goetsu}{IPSJ}[shiyoh@ruri.waseda.jp]
\author{Tetsuya Sakai}{IPSJ}[tetsuya@waseda.jp]

\begin{abstract}
  We report on an investigation into how different types of failures in
  a voice user interface (VUI) affects user frustration.
  To this end, we conducted a pilot user study ($n=10$) and a main user study ($n=30$), both
  with a simple voice-operated calendar application that we built using the Alexa Skills Kit.
  In our pilot study, we identified three major failure types
  as perceived by the users, namely,
  Reason Unknown, Speech Misrecognition, and Utterance Pattern Match Failure,
  along with more fine-grained failure types from the developer's viewpoint
  such as Intent Pattern Match Failure and
   Intent Misclassification.
  Then, in our main study,
  we set up three user tasks that were designed to each induce a specific failure type,
  and collected user frustration ratings for each task.
  Our main findings are:
  (a)~Users may be relatively tolerant to user-perceived Speech Misrecognition,
  and not so to user-perceived Reason Unknown and Utterance Mattern Match Failures;
  (b)~Regarding the relationship between developer-perceived and user-perceived failure types,
  68.8\% of
  developer-perceived
  Intent Misclassification instances caused user-perceived Reason Unkown failures.
  From (a) and (b),
  a practical design implication would be to try to prevent Intent Misclassification from happening
  by carefully crafting the utterance patterns for each intent.
\end{abstract}

\maketitle

\section{Introduction}
\label{s:intro}
Voice User Interfaces (VUIs) for dialogue systems have started to penetrate into our daily lives,
in the form of smart speakers such as Amazon Alexa and smart phones features such as Siri.
While many consumers regard these services as Artificial Intelligence
and may have various expectations for them,
current VUI interactions often fail for various trivial reasons.
Moreover,
the VUIs are generally not good at communicating
the reasons of failures to the user: when they fail to properly process the user's utterance for whatever reason,
they often just say ``\textit{I'm sorry, I don't understand}.'' or something similarly uninformative.
Hence, how the users perceive VUI failures may be different
from the actual failures as categorised from the developer's point of view.
We argue that
it is important to understand the different failure types from both developer and user perspectives,
how the two perspectives align,
and
how each failure type affects
user frustration,
so that
VUI and dialogue system developers can try to improve
on the failure types that matter most.
In the present study,
we primarily focus on the users' perception of VUI failures,
and investigate how the different failure types affect user frustration.

The present study consists of a pilot user study~\cite{uist2019} ($n=10$) and a main study ($n=30$);
both leveraged a simple voice-operated calendar application that we built for the purpose of this study
uusing the Alexa Skills Kit \footnote{\url{https://developer.amazon.com/en-US/alexa/alexa-skills-kit}}.
In the pilot study,
we identified major VUI failure types from both developer and user perspectives.
The failure types from the developer perspective are:
\begin{description}
\item[D1] Intent Pattern Match Failure;
\item[D2] Slot Value Extraction Failure;
\item[D3] System Not In Listen Mode;
\item[D4] Intent Misclassification;
\item[D5] Utterance Not Directed To System;
\item[D6] Partial Speech Misrecognition;
\item[D7] Complete Speech Misrecognition.
\end{description}
In contrast,
the user-perceived failure types that we identified through interviews are naturally more coarse-grained:
\begin{description}
\item[U1] Reason Unknown;
\item[U2] Speech Misrecognition;
\item[U3] Utterance Pattern Match Failure.
\end{description}
More explanations of these failure types will be given in Section \ref{s:pilot}.

The objective of our main study is to investigate
how the different failure types affect user frustration.
While we cannot directly control how our participants will perceive failures,
we gave them
three simple calendar manipulation tasks (voice-operated in Japanese) that were
designed to each
induce a specific developer-perceived failure type:
\begin{description}
\item[T1] Create a new event using a series of voice commands, by uttering the date and time
(\textit{December 25, 18:30-20:00}),
name of the event (\textit{Drinking party with part-time workers}), and the venue (\textit{Ginza Station}) separately (not designed specifically to induce any failures);
\item[T2] Create a new event by specifying all required information in one utterance (date and time: \textit{December 31, 9:00-11:00}, name: \textit{Study session}, venue: \textit{Takadanobaba Station})
 (designed to induce D1: Intent Pattern Match Failure,
since the utterance pattern for this intent requires
the user to specify all of these slot values in one go in a specific syntax);

\item[T3] Modify the name of the event created in T1 from \textit{Drinking party}
to \textit{Christmas}
(designed to induce D4: Intent Misclassification: we discovered in our pilot study that
Alexa automatically converts \textit{Christmas} into \textit{December 25},
which causes our VUI to misclassify the ``Modify Event Name'' intent).
\end{description}
As we shall explain later in Section \ref{s:main},
 T1 induced many U2 instances,
T2 induced many U3 instances, and
T3 induced many U1 instances,
and hence we managed to emphasise different user-perceived failures with the three different tasks.
Our main findings are:
\begin{description}
  \item[(a)]
Users may be relatively tolerant to what they perceive as speech recognition errors (U2),
  and not so when they do not understand the failures (U1),
  or when they feel that their wordings were not understood by the VUI (U3).
  \item[(b)]
Regarding the relationship between developer-perceived and user-perceived failure types,
  68.8\% of D4 (Intent Misclassification) instances caused U1 (Reason Unkown).
\end{description}

This paper concludes with a design implication based on (a) and (b).

\section{Related Work}

There is a body of research that tries to gain insight
into problems with VUIs and conversational agents (CAs)
by observing and/or interviewing users who use such systems regularly.
For example, Luger and Sellen~\cite{badPa}
interviewed $n=14$ regular users of conversational agents such as Siri
and concluded:
\textit{users had poor mental models of how their CA worked and that
these were reinforced through a lack of meaningful feedback regarding
system capability and intelligence.}
Porcheron \textit{et al.}~\cite{Everyday}
report on an analysis of audio data from
month-long deployments of Amazon Echo
and stress the importance
of system response design
as the design of interactional resources for users.
Pyae and Joelsson~\cite{investigatingUsability}
conducted a web-based survey
to which $n=114$ Google Home users responded;
the study lists up some problems that
the users encountered (according to the users' viewpoints),
such as
``Non-English words are not correctly captured by the device,''
``Commands have to be repeated to accomplish a task,''
and
``Multiple commands in a single transaction cannot be captured.''
Sciuto \textit{et al.}~\cite{HeyAlexa}
reported on a study of Amazon Alexa users
which involved both log analysis and in-home interviews.

In contrast to the above line of research that involves real users of commercial VUIs,
our study involves user studies in a controlled environment with a very simple VUI application;
the two approaches are clearly complementary.
Below, we also discuss some existing studies based on controlled studies.

The present study offers
VUI failure types from both the developer's and the user's perspectives,
as well as an analysis of how the two are related.
In previous work, there have also been a few studies
that listed up different failure types based on controlled studies.
For example,
in the context of conversational search,
Jiang, Jeng, and He~\cite{inputError} identified
the following voice input error types:
\textit{Speech Recognition Error},
\textit{System Interruption} (the participant was interrupted
before her voice command was complete),
and
\textit{Query Suggestion} (Google voice search generated a query
not uttered by the user).
The work of Myers \textit{et al.}~\cite{Patterns}, which was the direct inspiration of the present
study, identified the following four obstacles in VUIs through an experiment ($n=12$) with their calendar application called DiscoverCal:
\textit{Unfamilar Intent} (the VUI cannot parse the utterance for an existing intent, or
the participant tries to execute an intent not supported by the VUI),
\textit{NLP error} (the VUI maps the user utterance to an incorrect intent),
\textit{Failed Feedback} (for example,
the VUI did not make it clear to the user that
the date and time must be uttered in one go to make an entry into a calendar),
and
\textit{System Error} (e.g. bugs).
We note  that the above taxonomy is based on the developer's point of view,
i.e., what is really happening in the system internally.

In a follow-up study with $n=50$ participants,
Myers \textit{et al.}~\cite{Impact}
investigated the impact of user characteristics
on VUI task performance; they concluded
that while programming experience did not have a wide-spread impact on their performance metrics (e.g., total time spent on the tasks,
number of times the user had to repeat an intent, etc.)
assimilation bias (i.e., prejudice due to prior VUI experience) did.
They remark that, while Luger and Sellen~\cite{badPa}
reported that participants with more technical knowledge
were self-reported as being more patient and willing to utter more
to accomplish a VUI task,
their own results based on the total number of words uttered indicated otherwise.
Furthermore,
based on the above studies with DiscoverCal,
Myers~\cite{Adaptive} proposes
that the system's guidance to users should
differ according to the user's proficiency.

While we argue that our analysis of VUI failures from both
developer and user perspectives is a strength compared to prior art,
we acknowledge that
our participants are not representative of
the general consumers:
most of them are Computer Science (CS) students.
This limitation is also discussed in Section \ref{s:conclusions}.
To the best of our knowledge, however,
the present study is the first
to show that different VUI failure types
affect user frustration differently.

\section{Calendar Application}
\label{s:calendar}

Inspired by the work of Myers {\it et al.}~\cite{Patterns},
we built a voice-operated calendar application using Alexa Skills Kit
in order to conduct our pilot and main user studies.
Our application
consists of a VUI and a calendar GUI for visual feedback,
and can let the user create a new event on the calendar,
delete it, or modify it.
Internally,
the system maps a user's utterance to \emph{intents} (e.g., create an event, delete an event)
based on rule-based utterance patterns,
and then fills in \emph{slots} required in that intent wherever necessary
(e.g., name of the event, date and time of the event).
Whenever the system fails to process the user request,
it returns the generic message:
``Sorry, I cannot understand your request'' (in Japanese).

\section{Pilot Study}\label{s:pilot}

This section briefly describes our pilot study with $n=10$ participants~\cite{uist2019}.
Its main objective was to identify different failure types from both developer and user perspectives.
The user study design is similar to our main study,
so here we shall focus on the parts specific to the pilot study.

All of our pilot study participants were Japanese male students from the CS department of Waseda university;
 7 of them owned a smart speaker.
 Each participant was instructed to conduct four tasks:
 the first three were similar to T1-T3 for the main task;
 the fourth task was to \emph{delete} an existing event.
 (We obtained very few VUI failures from the delete task in the pilot study;
 hence it was dropped for the main study.)
 The instructions given to the participants were similar to those for the main study:
 see T1-T3 described earlier.
 All participants were asked to continue to try the task at least twice when they encountered a failure
 during eack task.

 After completing the tasks,
 the first author interviewed each participant in a face-to-face session;
 the interviews were recorded on a smartphone
 and later manually transcribed for analysis.
 On average, participants spent 11.0 minutes to complete the tasks,
and 9.1 minutes for the interview.
In the interviews, participants were asked how they felt and what
they thought about the failures they encountered during each task.
By manually analysing the actual dialogues and the interviews,
we identified 7 failure types from the developer perspective,
and 3 from the user perspective, as we have described in Section \ref{s:intro}.

Here, we briefly explain each failure type that we have identified.
As was mentioned in Section \ref{s:calendar},
our VUI application was developed by setting up several intents,
where each intent is associated with several predefined utterance patterns.
D1 means that the user's utterance did not match any of the utterance
patterns of any intent;
D2 means that the mapping to an intent was successful,
but that at least one required slot value could not be extracted from the user utterance;
D3 means that the user uttered a command when the VUI was not listening;
D4 means that the user's utterance was mapped to an incorrect intent;
D5 means that the user's utterance was not meant for the VUI (i.e., the user was talking to herself);
D6 means that the user's utterance was only partially successfully recognised
(e.g., when the user says ``from 9 to 11 o'clock'' and the system only recognises ``11 o'clock'');
D7 means complete speech recognition error.
On the other hand, the user's diagnosis of failures is naturally less specific.
U1 means that the user has no idea why the VUI fails to respond properly to the user's command;
U2 means that the user suspects that speech recognition
is the problem (and therefore a possible action she might take next
would be to repeat her previous utterance more slowly and clearly).
In contrast, when the user detects a U3,
this means that
she assumes that the VUI cannot accept the particular syntax of the uttered sentence
(and therefore possibly try to rephrase the same request).

\section{Main Study}\label{s:main}

Having identified the developer-perceived and user-perceived VUI failure types,
we proceeded with our main study ($n=30$),
where the objective was to investigate
how different failure types affect user frustration.
All of our participants were Japanese in their 20's and
 had a science background, with 23 with a CS background;
20 were male and 10 female;
27 were students at Waseda University
and the other three were recent graduates from the same university.
Regarding experience with smart speakers,
8 were regular users, 20 used them a few times before,
and 2 had no experience.
Each participant was asked to conduct tasks T1-T3 discussed in Section \ref{s:intro},
and then was interviewed by the first author after completing all three tasks
to see what types of failures were perceived during each task session.
On average, participants spent 9.1 minutes to complete the tasks,
and 10.3 minutes for the interview.
Thus, for each participant-task pair,
we analysed the VUI dialogues and the post-hoc interviews
to manually identify D1-D7 as well as U1-U3.
Moreover, each participant was asked to rate her overall frustration
for each of the three tasks on a Likert scale (1-5).

As was mentioned in Section \ref{s:intro},
it turned out that
T1 induced many U2 (Speech Misrecognition) instances,
T2 induced many U3 (Utterance Pattern Match Failure) instances,
and
T3 induced many U1 (Reason Unknown) instances.
Hence, even though we have the user frustration ratings at the task level
and not for each failure within the task\footnote{
User-perceived failures were identified by a post-hoc analysis of the interviews;
therefore, it was not possible for the participants to provide
a frustration rating for each failure  during the interviews.
},
we can shed some light on the relationship between
user-perceived failure types and the user frustration
by comparing the mean frustration scores across T1-T3.

Table~\ref{t:mean} shows the mean frustration (MF) scores over the $n=30$ participants
for each task.
Table~\ref{t:tukey} shows the results of the paired Tukey HSD (Honesty Significant Difference) test~\cite{sakai}:
it can be observed that
the MF for T2 and that for T3 are statistically significantly higher than that for T1,
while the difference between T3 and T2 is not statistically significant.

Table~\ref{t:U-vs-T} shows the distribution of user-perceived failures within each task of our main study.
The dominant user-perceived failure type for each task is shown in bold.
Since participants were more frustrated with T3 than with T1,
and U1 constitutes
$136/199=68.3$\% of the user-perceived failures in T3,
the results suggest that
U1 (Reason Unknown) may have a strong negative impact on user frustration.
Similarly, since participants were more frustrated with T2 than with T1,
and U3 constitutes
$101/178=56.7$\% of the user-perceived failures in T2,
the results suggest that
U3 (Utterance Pattern Match Failure) may also have a negative impact on user frustration.
In contrast, it appears that the participants were
relatively tolerant to what they perceive as
speech recognition errors (U2),
which constitutes
$48/104=46.2$\% of the user-perceived failures in T1.

\begin{table}[t]
\centering

\begin{tabular}{c|r}
\hline
Task	&Mean Frustration ($n=30$)\\
\hline
T1		&2.73\\
T2 		&3.97\\
T3		&4.40\\
\hline
\end{tabular}
\caption{Mean frustration scores for each task in the main study.}\label{t:mean}

\begin{tabular}{c|r|r|l}
\hline
           &diff        &simultaneous 95\%CI     &$p$-value\\
\hline
T2-T1 &1.233		&[0.599, 1.867]			&0.0000523\\
T3-T1 &1.667		&[1.033, 2.301]			&0.0000001\\
T3-T2 &0.433		&[$-$0.201, 1.067]			&0.236\\
\hline
\end{tabular}
\caption{Paired Tukey HSD test results for the differences in means shown in Table~\ref{t:mean}.
}\label{t:tukey}




\begin{tabular}{c|r|r|r|r}
\hline
		&T1			&T2			&T3			&total\\
\hline
U1		&26			&40 			&\textbf{136}	&202(42.0\%)\\
U2		&\textbf{48}	&37			&45			&130(27.0\%)\\
U3		&30			&\textbf{101}	&18			&149(31.0\%)\\
\hline
total		&104		&178		&199		&481(100\%)\\
\hline
\end{tabular}
\caption{Distribution of user-perceived failures within each task.
The most dominant failure type for each task is shown in bold.}\label{t:U-vs-T}

\end{table}


From the rightmost column of Table~\ref{t:U-vs-T},
it can be observed that, of the 481 user-perceived failures
across all tasks,
U1 constituted 42.0\%,
U3 constituted 31.0\%, and
U2 constituted 27.0\%.
In fact, from the developer's viewpoint,
there were more failures during the user experiments:
from the dialogues, we identified 586 developer-perceived failures,
which means that $(586-481)/586=17.9$\% of the actual
failures were not detected by the participants (or at least,
the participants did not mention them in their interviews).
Among the 586 developer-perceived failures, the dominant
types were D4 (39.9\%), D1 (29.2\%), D2 (13.1\%), D3 (10.6\%), and D7 (5.6\%).

\begin{figure}[t]
\begin{center}



\includegraphics[width=60mm]{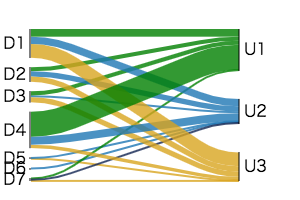}
\caption{Relationship between Developer-Perceived and User-Perceived Failure Types (based on user-perceived 481 failures).}\label{f:test201909}

\end{center}
\end{figure}

Figure~\ref{f:test201909} visualises
how the developer-perceived
failures
were perceived by the participants;
note that only the aforementioned 481 failure instances that were detected by the participants are shown here.
The most remarkable feature of this figure is that
D4 (Intent Misclassification) often translates to U1 (Reason Unknown);
more specifically,
of the 189 D4 instances observed across the tasks,
as many as 130 (68.8\%) were perceived by the participants as U1.
Since we have observed that U1
may have a strong negative impact on user frustration,
one practical approach to reducing user frustration in VUI applications
would be to try to minimise D4 incidents.
This can probably be achieved to some extent
by carefully crafting the utterance patterns for each intent.
However, note that the present study does \emph{not}
show that Intent Misclassification is the main cause of VUI failures:
recall that we intentionally induced  Intent Misclassifications just
to verify that different failure types affect user frustration differently.
All we claim is that  Intent Misclassification is something that VUI application developers should try to avoid.

The second strongest signal from Figure~\ref{f:test201909}
is that
D1 (Intent Pattern Match Failure) is often perceived by the participant as U3 (Utterance Pattern Match Failure):
of the 150 D1 instances observed,
72 (48.0\%) were perceived by the participants as U3.
That is, the diagnosis by the participants were correct in these cases,
although they may not necessarily be aware of the fact
that our VUI application
is composed of a set of intents
where each intent is associated with a set of utterance patterns.

\section{Conclusions and Future Work}
\label{s:conclusions}

Through our pilot study with a simple voice-operated calendar application,
we identified both developer-perceived and user-perceived failure types;
then, in our main study, we collected participants' frustration scores
for each of our three tasks (T1-T3) that were designed to induce specific failure types.
Our main findings are:
\begin{description}
\item[(a)] The mean frustration score of T1 (in which U2: Speech Misrecognition constituted 46.2\%)
was statistically significantly lower than that of
T2 (in which U3: Utterance Pattern Match Failure constituted 56.7\%) and
that of
T3 (in which U1: Reason Unknown constituted 68.3\%).
Hence,
users may be relatively tolerant to what they perceive as speech recognition errors (U2),
and not so when they do not understand the failures (U1),
or when they feel that their wordings were not understood by the VUI (U3).
\item[(b)] Regarding the relationship between developer-perceived and user-perceived failure types,
68.8\% of
D4 (Intent Misclassification) instances caused U1 (Reason Unkown).
\end{description}
One design implication based on the above two findings would be:
\emph{try to avoid Intent Misclassification,
since this very likely leads to failures for reason unknown from the user's point of view,
which in turn are likely to cause user frustration.}
For current rule-based VUI applications,
Intent Misclassifications can be suppressed to some extent
by carefully crafting the utterance patterns for each intent.
However, as we have pointed out earlier,
our study does \emph{not} show
that Intent Misclassification is the main cause of user frustration.

The participants we hired were mostly CS students in their 20's, and therefore arguably closer to developers than to general consumers
who have no knowledge of how VUI applications are implemented. Hence
our failure type taxonomies may not apply to them:
as an extreme situation, for some consumers, all failures may be of the Reason Unknown type.
Hence, as future work,
we would like to conduct a follow-up experiment that covers
a wider variety of user backgrounds.
Moreover, while the present study
collected user frustration scores at the task level,
we would like to explore nonintrusive ways to
collect frustration signals
for each user-perceived failure that occurs during each task,
so that
we can directly discuss the relationship between user-perceived
failure types and user frustration.
Furthermore, we would like to
establish a diagram  similar to Figure~\ref{f:test201909}
based on real VUI failure distributions
as opposed to our induced failures:
this
should be useful for
designing VUI applications that provide
informative failure responses for avoiding
or recovering from
dialogue breakdowns.



\begin{biography}
\profile{m}{情報 太郎}{1970年生．1992年情報処理大学理学部情報科学科卒．
1994年同大大学院修士課程了．同年情報処理学会入社．オンライン出版の研究
に従事．電子情報通信学会，IEEE，ACM 各会員}
\profile{n}{処理 花子}{1960年生．1982年情報処理大学理学部情報科学科卒．
1984年同大大学院修士課程了．1987年同博士課程了．理学博士．1987年情報処
理大学助手．1992年架空大学助教授．1997年同大教授．オンライン出版の研究
に従事．2010年情報処理記念賞受賞．電子情報通信学会，IEEE，IEEE-CS，ACM
各会員}
\profile{s}{学会 次郎}{1950年生．1974年架空大学大学院修士課程了．
1987年同博士課程了．工学博士．1977年架空大学助手．1992年情報処理大学助
教授．1987年同大教授．2000年から情報処理学会顧問．オンライン出版の研究
に従事．2010年情報処理記念賞受賞．情報処理学会理事．電子情報通信学会，
IEEE，IEEE-CS，ACM 各会員}
\end{biography}
\bibliographystyle{ipsjsort}
\bibliography{tech-jsample}

\end{document}